# Morphological Evolution and Surface/Interface Fe-oxide formation in Epitaxial Fe(001)/MgO(001) Thin Film

### Md. Shahid Jamal and Dileep Kumar*

UGC-DAE Consortium for Scientific Research, Khandwa Road, Indore-452001, India
*Corresponding author: dkumar@csr.res.in*

### Abstract

Epitaxial Fe film has been grown on ion beam sputtered MgO(001) substrate by electron beam evaporation. Reflection high energy electron diffraction (RHEED) and transport measurements (TM) were performed simultaneously during the growth of the film. While TM provided information about film morphology, RHEED provided information about the structure of the film. The film grows via the Volmer–Weber mechanism, where epitaxial islands grow larger to impinge with other islands and eventually coalesce into a continuous film at around 1.8 nm thickness. Unlike what has been reported in the literature, there is no sign of the iron oxide layer forming at the Fe/MgO interface, not even at 300 °C. A high-quality epitaxial $Fe_3O_4$ layer is found to develop on the Fe surface when the sample is annealed under a moderate oxygen partial pressure.

## Introduction

Over a decade Fe/MgO system has been the subject of extensive investigations due to numerous interesting phenomena such as tunnel magnetoresistance [1], perpendicular magnetic anisotropy [2], interlayer coupling [3], and the exchange bias [4]. Here, interface oxidation is well known for drastically altering the behavior of the system. The oxide layer formation at the Fe/MgO interface and its clear role in modifying magnetic and transport properties is still missing in the literature.

The present work deals with the growth of an epitaxial Fe film on the MgO(001) substrate, where real-time characterization was done simultaneously using reflection high energy electron diffraction (RHEED) and transport measurements (TM) measurements during and after the film growth. TM provided information about the different stages of the film growth, whereas RHEED gave information about the structure and morphology of the film surface.

## Experimental

Fe film of ~3.0 nm thickness is grown on MgO(001) substrate using e-beam evaporation and characterized in-situ using RHEED and TM during growth. Au contact pad has been deposited (~ 30 nm) on the substrate for TM. Before Fe deposition, the substrate was cleaned by annealing at 500 °C for 30 min followed by Ar+ ion etching at 1 KeV for 20 min. In-situ RHEED measurements were also performed after annealing Fe film up to 300 °C. All the RHEED images were collected along [100] direction of Fe. For magnetic properties, magneto-optic Kerr effect (MOKE) measurements were performed before and after annealing at 300 °C (with and without oxygen partial pressure).

## Results and Discussion

Fig. 1(a) shows a schematic, where the direction of RHEED, deposition, ion gun, and TM are mentioned. Total sheet resistance ($R_{Fe}$) vs Fe film thickness ($d_{Fe}$) is plotted in Fig. 1(b) and divided into three different regions (A, B, and C) to understand different stages of Fe growth. $R_{Fe}$ in the region "A" is large (~19 k/□) and remains almost the same up to a film thickness of ~0.8 nm due to the small island state of Fe film. In this state, island to island hopping probability of electrons is negligible, and therefore, resistance is large. In this region, the resistance is mainly coming from the substrate. With increasing $d_{Fe}$, beyond ~0.8 nm (region B), isolated islands start to connect, and hence, resistance decreases drastically.

After ~1.8 nm thickness (region C), it decreases slowly due to the formation of a continuous layer. RHEED images 1(c-e) correspond to 0.53 nm, 0.92 nm, and 3.0 nm thick Fe films. The spotty and streaky pattern in all images indicates the epitaxial growth of Fe (001) on the MgO(001) substrate [5]. As RHEED is a surface-sensitive technique up to ~ 3 nm, therefore all three images are expected to have structural information throughout the Fe thickness (including interface). It may be noted that the absence of any other additional line (other than Fe) clearly shows the absence of an oxide layer at the surface and interface of the Fe/MgO layer.

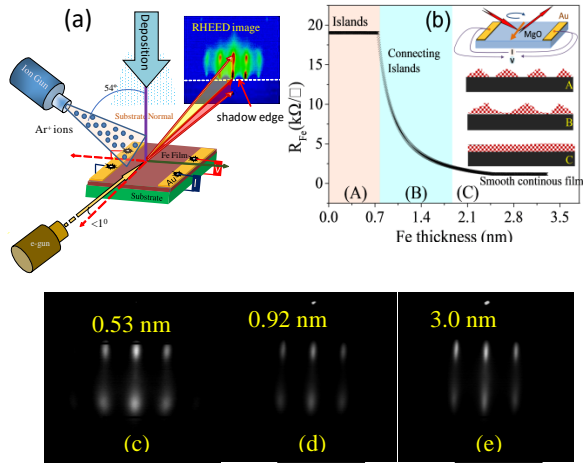

**Fig. 1.** (a) Schematic of the in-situ measurements.(b) Sheet resistance ($R_{Fe}$) Vs Fe thickness on MgO(001) substrate and morphological evolution; A- isolated islands, B- connecting islands, and C- continuous film. Insets show Au contacts for TM measurement. RHEED images at Fe layer thickness; (c) 0.53 nm, (d) 0.92 nm, and (e) 3.0 nm.

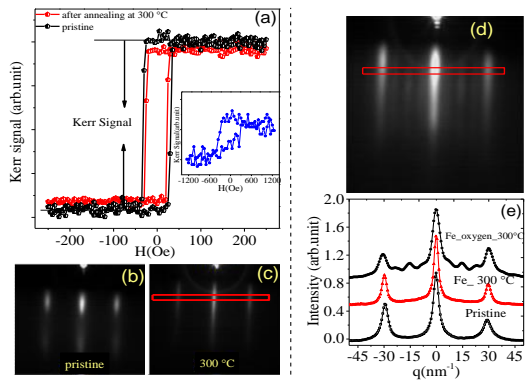

**Fig. 2**(a,) & (b, c) show hysteresis loops and RHEED images of 3 nm thick Fe film before and after annealing (300 °C), respectively. Loop (inset of Fig. 2a) and RHEED (Fig. 2d) were taken after annealing at 300 °C in oxygen partial pressure of 5×10⁻⁷ mbar. (e) Intensity profile of all RHEED images.

To further study the effect of annealing, Fe film is heated at 300 °C temperature for 20 min. RHEED and longitudinal MOKE measurements [6, 7], before and after annealing, are shown in Fig. 2. Hysteresis loops (Fig. 2a) and RHEED (Fig. 2b & c) patterns are hardlay affected by the annealing. A slight difference in the loop signal (height) and coercivity (Hc), is mainly due to the structural relaxation after heating. The same has also been confirmed by the shaping of the RHEED lines after annealing (Fig. 2c). On the other hand, when the oxygen partial pressure of 5×10⁻⁷ mbar is created in the chamber for 5 min during heating at 300 °C, a drastic change in the loop signal (inset of Fig. 2a) and presence of additional lines between Fe lines in RHEED image (Fig. 2d) can be understood in term of the surface oxidation of the Fe film. The intensity line profile (along the marked rectangular line in Fig. 2c & 2d) of all three images is extracted and shown in Fig. 2e. The appearance of three additional lines in between two intense high lines corresponds to the $Fe_3O_4$ [8].

Drastic reduction in the Kerr signal after the formation of the $Fe_3O_4$ layer on the Fe surface can be understood by the expression of Kerr intensity in the longitudinal geometry of MOKE, where the Kerr signal is linearly related to the thickness ($d_{Fe}$) of magnetic film in the ultrathin regime and given by the relation [9, 10]-

$$\varphi_{long} = \frac{4\pi n_{sub} Q d_{Fe} \theta}{\lambda (1 - n_{sub}^2)}$$

Q is the magneto-optical constant and θ is the angle (~45°) of incidence measured from surface normal. As the oxide formation took place at the cost of the Fe layer on the surface, therefore the reduction in the Kerr signal could be attributed to the decrease in $d_{Fe}$. On the other hand, an increase in the Hc may be due to the increase in domain wall pinning at the $Fe_3O_4$/Fe interface.

## Conclusion

Fe film is grown epitaxially on MgO(001) substrate to study morphological evolution and surface/interface Fe-oxide formation. Systematic studies of epitaxial Fe film were done in-situ during different stages of the Fe growth using RHEED, MOKE, and TM. Fe film is found to grow via Volmer–Weber mechanism, where epitaxial islands grow larger to impinge with other islands and eventually coalesce into a continuous film at around 1.8 nm thickness. Unlike the literature [11], no evidence of the formation of an iron oxide layer is observed at the Fe/MgO interface even up to thermal annealing at a temperature of 300 °C. A well-defined epitaxial $Fe_3O_4$ layer is found to form on the Fe(001)/MgO(001) surface when the sample is annealed at 300 °C for 5 min in the presence of oxygen partial pressure.